\documentclass[12pt,a4paper]{article}
\usepackage{amsfonts,latexsym}
\usepackage{amsmath,amssymb}
\usepackage{graphicx,color}

\renewcommand{\thefootnote}{\fnsymbol{footnote}}

\begin{document}

\vspace{12mm}

\begin{center}
{{{\Large {\bf Einstein-singleton theory and its power spectra \\in de Sitter inflation}}}}\\[10mm]

{Yun Soo Myung$^a$\footnote{e-mail address: ysmyung@inje.ac.kr}, Taeyoon Moon$^a$\footnote{e-mail address: tymoon@inje.ac.kr}, and Young-Jai Park$^b$\footnote{e-mail address: yjpark@sogang.ac.kr}}\\[8mm]

{$^a$Institute of Basic Sciences and Department  of Computer Simulation, Inje University Gimhae 621-749, Korea\\[0pt]}
{$^b$Department of Physics, Sogang University, Seoul 121-742, Korea\\[0pt]}

\end{center}
\vspace{2mm}

\begin{abstract}
We study the Einstein-singleton theory during de Sitter inflation
since  it provides a way of  degenerate fourth-order scalar theory.
We obtain an exact solution expressed in terms of the
exponential-integral function  by solving the degenerate
fourth-order scalar equation in de Sitter spacetime. Furthermore, we
find that its power spectrum blows negatively up in the superhorizon
limit, while it is negatively scale-invariant in the subhorizon
limit. This suggests that the Einstein-singleton theory contains the
ghost-instability and thus, it is not  suitable for developing   a
slow-roll inflation model.

\end{abstract}
\vspace{5mm}

{\footnotesize ~~~~PACS numbers: 04.50.Kd,98.80.Cq,98.80.Jk }

{\footnotesize ~~~~Keywords: singleton, power spectrum, de Sitter
inflation}

\vspace{1.5cm}

\hspace{11.5cm}{Typeset Using \LaTeX}
\newpage
\renewcommand{\thefootnote}{\arabic{footnote}}
\setcounter{footnote}{0}

%%%% Introduction %%%%

\section{Introduction}

The  single-field inflation  is still known to be a  promising model
for describing the slow-roll (quasi-de Sitter)
inflation~\cite{Ade:2015lrj} when one chooses an appropriate
potential like the Starobinsky potential which originates from
$f(R)=R+R^2$ gravity~\cite{Starobinsky:1980te}. This Einstein-scalar
theory corresponds to a second-order tensor-scalar theory.

Our next question is to consider an Einstein-(higher-order) scalar
theory even though one may worry about a ghost state. For this
purpose, it was interesting to compute the power spectrum of a
massive singleton (other than inflaton) generated during de Sitter
(dS) inflation because its equation belongs to a fourth-order
equation. In order to compute the power spectrum, one has  to choose
the Bunch-Davies vacuum in the subhorizon limit of $z\to \infty$. In
addition, one needs to quantize the singleton canonically as the
inflaton did. However, it is hard  to obtain  a fully exact solution
to the fourth-order equation in dS spacetime. Instead, the authors
in ~\cite{Kehagias:2012pd} have investigated  the massive singleton
to show the dS/LCFT correspondence in the superhorizon limit of
$z\to 0$ as an extension to the dS/CFT correspondence. Recently, two
of us
 have shown  that the momentum correlators of LCFT take the
same form as the power spectra $\times k^3$ in the superhorizon
limit~\cite{Myung:2014pza}. This might  show that the dS/LCFT
correspondence works  for obtaining the power spectra in the
superhorizon limit.     Nevertheless, the limitation of these works
is that their computations are valid only in the superhorizon limit
because of difficulty in solving a fourth-order differential
equation in whole range $z$.

In this work we obtain an  exact solution and compute a complete
power spectrum of singleton by solving the degenerate fourth-order
scalar equation, which describes a propagation of a massless
singleton during dS inflation and by requiring the Pais-Uhlenbeck
quantization scheme for a degenerate fourth-order
oscillator~\cite{Pais:1950za,Mannheim:2004qz,Kim:2013mfa}. It turns
out that the singleton power spectrum blows negatively up in the
superhorizon limit, while it is negatively scale-invariant in the
subhorizon limit.  This suggests that the Einstein-singleton theory
is not a  candidate for a slow-roll inflation  because its power
spectrum might show ghost-instability.

%%%%%%%%%%%%%%%%%%%%%%%%%%%%%%%%%%%%%%%%%%%%%%%%%%%%%%%%%%%%%%%%%%%%%
%%%%%%%%%%%%%%%%%%%%%%%%%%%%%%%%%%%%%%%%%%%%%%%%%%%%%%%%%%%%%%%%%%%%%
%%%%%%%%%%%%%%%%%%%%%%%%%%%%%%%%%%%%%%%%%%%%%%%%%%%%%%%%%%%%%%%%%%%%%
\section{Einstein-singleton theory }
%%%%%%%%%%%%%%%%%%%%%%%%%%%%%%%%%%%%%%%%%%%%%%%%%%%%%%%%%%%%%%%%%%%%%
%%%%%%%%%%%%%%%%%%%%%%%%%%%%%%%%%%%%%%%%%%%%%%%%%%%%%%%%%%%%%%%%%%%%%
%%%%%%%%%%%%%%%%%%%%%%%%%%%%%%%%%%%%%%%%%%%%%%%%%%%%%%%%%%%%%%%%%%%%%

We introduce  the Einstein-singleton theory  where a dipole ghost
pair $\phi_1$ and $\phi_2$ are minimally coupled  to Einstein
gravity. The starting action  is a second-order scalar-tensor theory
given by
\begin{equation} \label{SGA}
S_{\rm ES}=S_{\rm E}+S_{\rm S}=\int d^4x
\sqrt{-g}\Big[\Big(\frac{R}{2\kappa}-2\Lambda\Big)-\Big(\partial_\mu\phi_1\partial^\mu\phi_2+\frac{\mu}{2}\phi_1^2
\Big)\Big],
\end{equation}
where $S_{\rm E}$  is introduced to feed  the dS inflation with
$\Lambda>0$ and $S_{\rm S}$ represents the singleton theory composed
of two scalars $\phi_1$ and
$\phi_2$~\cite{Flato:1986uh,Ghezelbash:1998rj,Kogan:1999bn,Myung:1999nd,Grumiller:2013at}.
Here we have $\kappa=8\pi G=1/M^2_{\rm P}$ with  the reduced Planck
mass $M_{\rm P}$ and $\mu$ is a coupling parameter.

After the metric variation, the Einstein equation is given by
\begin{equation} \label{ein-eq}
G_{\mu\nu} +\kappa \Lambda g_{\mu\nu}=\kappa T_{\mu\nu}
\end{equation}
with the energy-momentum tensor
\begin{equation}
T_{\mu\nu}=2\partial_{\mu}\phi_1\partial_\nu
\phi_2-g_{\mu\nu}\Big(\partial_\mu\phi_1\partial^\mu\phi_2+\frac{\mu}{2}\phi_1^2\Big).\end{equation}
Importantly, two scalar fields  are coupled to be
\begin{equation} \label{b-eq1}
\nabla^2\phi_1=0,~~\nabla^2\phi_2=\mu\phi_1,
\end{equation}
which lead to  a degenerate fourth-order   equation
\begin{equation} \label{b-eq2}
\nabla^4\phi_2=0.
\end{equation}
It can describe a fourth-order scalar theory  because $S_{\rm S}$
reduces  to  the fourth-order scalar theory when eliminating  an
auxiliary field $\phi_1$ as~\cite{Rivelles:2003jd}
\begin{equation} \label{singleton}
S^4_{\rm S}=\frac{1}{2\mu} \int
d^4x\sqrt{-g}\nabla^2\phi_2\nabla^2\phi_2,
\end{equation}
which provides (\ref{b-eq2}) directly. Choosing  the
 vanishing scalars,  the solution of dS spacetime comes out as
 \begin{equation}
 \bar{R}=4\kappa \Lambda,~~\bar{\phi}_1=\bar{\phi}_2=0.
 \end{equation}
Explicitly,  dS-curvature quantities are given by
\begin{equation}
\bar{R}_{\mu\nu\rho\sigma}=H^2(\bar{g}_{\mu\rho}\bar{g}_{\nu\sigma}-\bar{g}_{\mu\sigma}\bar{g}_{\nu\rho}),~~\bar{R}_{\mu\nu}=3H^2\bar{g}_{\mu\nu}
\end{equation}
with a Hubble parameter $H=\sqrt{\kappa \Lambda/3}$. We select  the
dS background explicitly by choosing a conformal time $\eta$
\begin{eqnarray} \label{frw}
ds^2_{\rm dS}=\bar{g}_{\mu\nu}dx^\mu
dx^\nu=a(\eta)^2[-d\eta^2+\delta_{ij}dx^idx^j],
\end{eqnarray}
where the conformal and cosmic  scale factors are given by
\begin{eqnarray}
a(\eta)=-\frac{1}{H\eta},~ a(t)=e^{Ht}.
\end{eqnarray}
 During the dS inflation, $a(\eta)$ goes from small to a very
large value like $a_f/a_i\simeq 10^{30}$, which corresponds to the
fact  that the conformal time $\eta=-1/a(\eta)H$ runs from $-\infty$
(subhorizon) to ${}^{-}0$ (superhorizon). The Penrose diagram is
depicted in Fig. 1.  Conformal invariance in $\mathbb{R}^3$ at
$\eta=-\epsilon$ is connected to the isometry group SO(1,4) of dS
space. In this case, the dS isometry group acts as conformal group
when fluctuations are superhorizon~\cite{Kehagias:2012pd}. Hence,
correlators are expected to be constrained  by conformal invariance.
Actually, a slice ($\mathbb{R}^3$) at $\eta=-\epsilon$ is employed
to calculate the power spectrum in the superhorizon limit.  On the
other hand, one introduces the Bunch-Davies vacuum to compute the
power spectrum in the subhorizon limit of $\eta\to -\infty$.
%%%%%%%%%%%%%%%%%%%%%%%%%%%%
\begin{figure*}[t!]
\centering
\includegraphics[width=.5\linewidth,origin=tl]{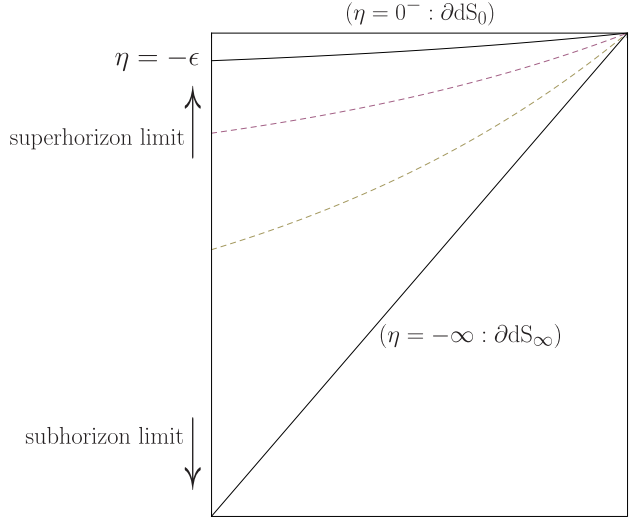}
\caption{Penrose diagram of dS spacetime with the UV/IR boundaries
(${\rm
\partial dS}_{\infty/0}$) located   at
$\eta=-\infty$ and  $\eta={}^{-}0$. A slice ($\mathbb{R}^3$)  near
$\eta=-\infty$ is introduced to compute the power spectrum in the
subhorizon limit, while a  slice ($\mathbb{R}^3$) at
$\eta=-\epsilon$ is employed to calculate the power spectrum in the
superhorizon limit. }
\end{figure*}
%%%%%%%%%%%%%%%%%%%%%%%%%%%%

We wish to choose  the Newtonian gauge of $B=E=0 $ and $\bar{E}_i=0$
for cosmological perturbation around the dS background (\ref{frw}).
In this case, the cosmologically perturbed metric can be simplified
to be
\begin{eqnarray} \label{so3-met}
ds^2=a(\eta)^2\Big[-(1+2\Psi)d\eta^2+2\Psi_i d\eta
dx^{i}+\Big\{(1+2\Phi)\delta_{ij}+h_{ij}\Big\}dx^idx^j\Big]
\end{eqnarray}
with transverse-traceless tensor $\partial_ih^{ij}=h=0$. Furtherore,
two scalar perturbations are defined by  \begin{equation} \phi_1=
0+\varphi_1,~~\phi_2= 0+\varphi_2.
\end{equation}
In order to obtain  the perturbed Einstein equations, one can
linearize the Einstein equation (\ref{ein-eq})  directly around the
dS spacetime as
\begin{eqnarray}
\delta R_{\mu\nu}(h)-3H^2h_{\mu\nu}=0 \to
\bar{\nabla}^2h_{ij}=0,\label{heq}
\end{eqnarray}
which describes a massless gravitational wave propagation.
Concerning  two-metric  scalars $\Psi$ and $\Phi$, their linearized
Einstein equations imply  that they are not physically propagating
modes. In addition,  we note that  there is no coupling between
$\{\Psi,\Phi\}$ and $\{\varphi_1,\varphi_2\}$ because of
$\bar{\phi}_1=\bar{\phi}_2=0$ in dS inflation. The vector $\Psi_i$
is also a non-propagating mode since it has no kinetic term. The
relevant linearized equations are those for two scalars
\begin{eqnarray} \label{singg-eq1}
&&\bar{\nabla}^2\varphi_1=0,\\
&&\label{sing-eq1}\bar{\nabla}^2\varphi_2=\mu\varphi_1,
\end{eqnarray}
which  are combined to provide a degenerate fourth-order scalar
equation
\begin{equation} \label{sing-eq2}
\bar{\nabla}^4\varphi_2=0.
\end{equation}
This  is our main equation to be solved to obtain the power spectrum
of a massless singleton during dS-inflation.

It seems appropriate to comment  that
Eqs.(\ref{singg-eq1})-(\ref{sing-eq2}) are different from those of a
massive singleton in~\cite{Myung:2014pza}:
$(\bar{\nabla}^2-m^2)\varphi_1=0,~(\bar{\nabla}^2-m^2)\varphi_2=\mu\varphi_1,~
(\bar{\nabla}^2-m^2)^2\varphi_2=0$. We could not solve the massive
singleton equation in the whole range of $\eta\in[-\infty,{}^{-}0]$.

%%%%%%%%%%%%%%%%%%%%%%%%%%%%%%%%%%%%%%%%%%%%%%%%%%%%%%%%%%%%%%%%%%%%%
%%%%%%%%%%%%%%%%%%%%%%%%%%%%%%%%%%%%%%%%%%%%%%%%%%%%%%%%%%%%%%%%%%%%%
%%%%%%%%%%%%%%%%%%%%%%%%%%%%%%%%%%%%%%%%%%%%%%%%%%%%%%%%%%%%%%%%%%%%%
\section{Propagation of massless singleton}
%%%%%%%%%%%%%%%%%%%%%%%%%%%%%%%%%%%%%%%%%%%%%%%%%%%%%%%%%%%%%%%%%%%%%
%%%%%%%%%%%%%%%%%%%%%%%%%%%%%%%%%%%%%%%%%%%%%%%%%%%%%%%%%%%%%%%%%%%%%
%%%%%%%%%%%%%%%%%%%%%%%%%%%%%%%%%%%%%%%%%%%%%%%%%%%%%%%%%%%%%%%%%%%%%

In order to compute the complete power spectrum, we have to know the
solution to singleton equations (\ref{sing-eq1}) and
(\ref{sing-eq2}) in the whole range of $\eta\in[-\infty,{}^{-}0]$.
For this purpose, two scalars $\varphi_{i}$ can be expanded in
Fourier modes $\phi^{i}_{\bf k}(\eta)$
\begin{eqnarray}\label{scafou}
\varphi_{i}(\eta,{\bf x})=\frac{1}{(2\pi)^{\frac{3}{2}}}\int d^3{\bf
k}~\phi^{i}_{\bf k}(\eta)e^{i{\bf k}\cdot{\bf x}}.
\end{eqnarray}
Eq.(\ref{sing-eq1}) leads to
\begin{eqnarray}\label{scalar-eq2}
\Bigg[\frac{d^2}{d \eta^2}-\frac{2}{\eta}\frac{d}{d
\eta}+k^2\Bigg]\phi^1_{\bf k}(\eta)=0.
\end{eqnarray}
Introducing a new variable $z=-k\eta$, Eq.(\ref{scalar-eq2}) can be
rewritten as
\begin{equation}\label{scalar-eq6}
\Big[\frac{d^2}{dz^2}-\frac{2}{z}\frac{d}{dz}+1\Big]\phi^{1}_{\bf
k}(z)=0
\end{equation}
whose  positive-frequency  solution with the normalization
$1/\sqrt{2k}$ is given by
\begin{equation} \label{scalar-eq7}
\phi^{1}_{\bf k}(z)=\frac{H}{\sqrt{2k^3}}(i+z)e^{iz}.
\end{equation}
This is the  typical  solution of a massless scalar propagating on
dS spacetime.

On the other hand,  plugging (\ref{scafou}) into (\ref{sing-eq2})
leads to the  fourth-order scalar equation
\begin{eqnarray}
\Big[\eta^2\frac{d^2}{d\eta^2}-2\eta\frac{d}{d\eta}+k^2\eta^2\Big]^2\phi^2_{\bf
k}(\eta)=0.\label{s2-eq2}
\end{eqnarray}
This equation  can be expressed in terms of $z$ as
\begin{equation}\label{sub-eq1}
\Bigg[\frac{d^4}{dz^4}+2\Big(1-\frac{1}{z^2}\Big)\frac{d^2}{dz^2}+\frac{4}{z^3}\frac{d}{dz}+\Big(1-\frac{2}{z^2}\Big)\Bigg]\phi^2_{\bf
k}=0
\end{equation}
whose full  solution is found to be
\begin{eqnarray} \label{sub-sol}
\phi^{2}_{\bf
k}(z)=\Big[\tilde{c}_2(i+z)+\tilde{c}_1\Big\{2i+(z-i)e^{-2iz}{\rm
Ei}(2iz)\Big\}\Big]e^{iz}
\end{eqnarray}
with two complex coefficients $\tilde{c}_1$ and $\tilde{c}_2$. This
is one of  our main results which states that the solution
(\ref{sub-sol}) is an exact solution to the fourth-order equation
(\ref{sing-eq2}).
 The c.c. of
$\phi^{2}_{\bf k}$ is  also a solution to (\ref{sub-eq1}). Here,
${\rm Ei}(2iz)$ is the exponential-integral function of a purely
imaginary number defined by~\cite{AS}
\begin{eqnarray}\label{eid}
{\rm Ei}(2iz)={\rm Ci}(2z)+i{\rm Si}(2z)-i\frac{\pi}{2},
\end{eqnarray}
where the cosine-integral and sine-integral functions are given by
\begin{eqnarray}
{\rm Ci}(2z)&=&-\int^{\infty}_{2z}\frac{{\cos}
t}{t}dt\longrightarrow\left\{\begin{array}{ll} z\to0:
\gamma+\ln[2z]+\Sigma^{\infty}_{k=1}\frac{(-1)^k(2z)^{2k}}{2k(2k)!} \label{cia}\\
z\to\infty: \frac{{\rm sin}(2z)}{2z}+{\cal O}\frac{1}{z^2}
\end{array}\right.,\\
&&\nonumber\\
{\rm Si}(2z)&=&\int^{2z}_{0}\frac{{\rm sin}
t}{t}dt\longrightarrow\left\{\begin{array}{ll}z\to0:
\Sigma^{\infty}_{k=1}\frac{(-1)^{k-1}(2z)^{2k-1}}{(2k-1)(2k-1)!} \label{sia}\\
z\to\infty:-\frac{{\cos}(2z)}{2z}+\frac{\pi}{2}+{\cal
O}\frac{1}{z^2}
\end{array}\right.
\end{eqnarray}
with the  Euler's constant $\gamma=0.577$ . Their behaviors are
depicted in Fig. 2.
%%%%%%%%%%%%%%%%%%%%%%%%%%%%%%%%%%%%%%%%%%%%%%%%%%%%%%%%%%%%%%%%%
\begin{figure}[t!]
\begin{center}
\begin{tabular}{cc}
\includegraphics[width=.8
\linewidth,origin=tl]{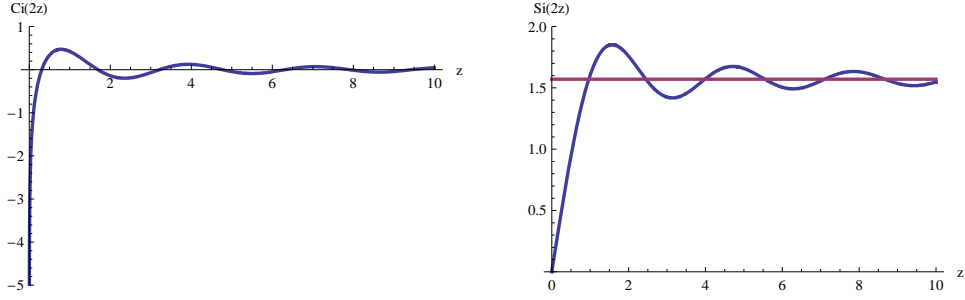}
\end{tabular}
\end{center}
\caption{Cosine-integral and Sine-integral functions as functions of
$z$. In the super horizon limit of $z\to0$, one finds  that
Ci$[2z]\to \gamma+\ln[2z]$ and Si$[2z]\to 0$. On the other hand, one
finds that Ci$[2z] \to \frac{{\rm sin}[2z]}{2z}$ and  Si$[2z]\to
\frac{\pi}{2}-\frac{{\rm cos}[2z]}{2z}$ in the subhorizon limit of
$z\to\infty$.}
\end{figure}
%%%%%%%%%%%%%%%%%%%%%%%%%%%%%%%%%%%%%%%%%%%%%%%%%%%%%%%%%%%%%%%%%
We note that ${\rm Ei}(2iz)$ satisfies the fourth-order equation
\begin{eqnarray} (z-i)z^3\frac{d^4{\rm Ei}(2iz)}{dz^4}&-&4iz^4 \frac{d^3{\rm Ei}(2iz)}{dz^3}+2z(i-z-4iz^2-2z^3)\frac{d^2{\rm Ei}(2iz)}{dz^2}\nonumber
\\
&-&4(i-z-iz^2+2z^3)\frac{d{\rm Ei}(2iz)}{dz}=8e^{2iz}
\end{eqnarray}
and its asymptotic behaviors are given by
\begin{eqnarray}\label{eiaf}
{\rm Ei}(2iz)\longrightarrow\left\{\begin{array}{ll} z\to0:
\gamma+\ln[2z]-\frac{i\pi}{2} \label{eia}\\
z\to\infty: -\Big[\frac{i}{2z}+\frac{1}{(2z)^2}\Big]e^{2iz}
\end{array}\right.
\end{eqnarray}
obtained from (\ref{eid}) together with (\ref{cia}) and (\ref{sia}).

 It is worth to point out
that the solution (\ref{sub-sol}) is suitable for choosing the
Bunch-Davies vacuum to give quantum fluctuations because  it shows
\begin{eqnarray}\label{phi2if}
\phi^{2}_{\bf
k}(z)\to_{z\to\infty}\Big[\Big(\tilde{c}_2+\frac{3}{2}\tilde{c}_1\Big)i+\tilde{c}_2
z\Big]e^{iz}.\end{eqnarray}
 Then, Eq.(\ref{sub-eq1}) in the subhorizon limit of $z\to \infty$
reduces to a degenerate fourth-order equation  which appeared  in
conformal gravity~\cite{Myung:2015vya}
\begin{equation} \label{phi2t}
\Big[\frac{d^2}{dz^2}+1\Big]^2\phi^{2}_{{\bf k},\infty}(z)=0
\end{equation}
whose solution is given by
\begin{equation} \label{phi2s-sol} \phi^{2}_{{\bf
k},\infty}(z)=(c'_1+c'_2z)e^{iz}.
\end{equation}
We note that after redefining $\tilde{c}_1$ and ${\tilde c}_2$,
Eq.(\ref{phi2if}) leads to Eq.(\ref{phi2s-sol}). The undetermined
constants $c'_1$ and $c'_2$ shows a feature of solution to the
fourth-order equation (\ref{phi2t}) when one compares these with the
fixed solution (\ref{scalar-eq7}) to the second order equation.

On the other hand, in the superhorizon limit of $z\to 0$,
Eq.(\ref{sub-eq1}) reduces to
\begin{equation}\label{super-eq10}
\Bigg[\frac{d^4}{dz^4}-\frac{2}{z^2}\frac{d^2}{dz^2}+\frac{4}{z^3}\frac{d}{dz}\Bigg]\phi^{2}_{{\bf
k},0}=0,
\end{equation}
whose solution is given by
\begin{equation}\label{solsuper}
\phi^{2}_{{\bf k},0}=\bar{c}_1+\bar{c}_2\ln[2z]
\end{equation}
with arbitrary constants $\bar{c}_1$ and $\bar{c}_2$. Especially,
the presence of $\ln[2z]$ dictates that (\ref{solsuper}) is the
solution to the fourth-order equation (\ref{super-eq10}).  In
deriving Eq.(\ref{super-eq10}) from Eq.(\ref{sub-eq1}), we neglect
the last term of $-\frac{2}{z^2}$ because it is subdominant in the
limit of $z\to 0$. We note that the full solution (\ref{sub-sol})
reduces  to Eq.(\ref{solsuper}) in the limit of $z\to0$ :
\begin{equation}
\phi^{2}_{{\bf
k},0}=i\Big[\tilde{c}_2+(2-\gamma+i\frac{\pi}{2})\tilde{c}_1\Big]-i\tilde{c}_1\ln[2z],
\end{equation}
when choosing
\begin{eqnarray}
\bar{c}_2=-i\tilde{c}_1,~~~~~\bar{c}_1=i\Big[\tilde{c}_2+\Big(2-\gamma+\frac{i\pi}{2}\Big)\tilde{c}_1\Big].
\end{eqnarray}

 Finally, we may determine one  coefficient $\tilde{c}_1$ by
making use of  Eq.(\ref{sing-eq1}) together with
Eqs.(\ref{scalar-eq7}) and (\ref{sub-sol}):
\begin{equation} \label{c1mu}
\tilde{c}_1=-\frac{\mu}{3H\sqrt{2k^3}}.
\end{equation}
However,  $\tilde{c}_2$ remains undetermined, but it will be
determined by the Wronskian condition in the next section.

%%%%%%%%%%%%%%%%%%%%%%%%%%%%%%%%%%%%%%%%%%%%%%%%%%%%%%%%%%%%%%%%%%%%%
%%%%%%%%%%%%%%%%%%%%%%%%%%%%%%%%%%%%%%%%%%%%%%%%%%%%%%%%%%%%%%%%%%%%%
%%%%%%%%%%%%%%%%%%%%%%%%%%%%%%%%%%%%%%%%%%%%%%%%%%%%%%%%%%%%%%%%%%%%%
\section{Power spectra}
%%%%%%%%%%%%%%%%%%%%%%%%%%%%%%%%%%%%%%%%%%%%%%%%%%%%%%%%%%%%%%%%%%%%%
%%%%%%%%%%%%%%%%%%%%%%%%%%%%%%%%%%%%%%%%%%%%%%%%%%%%%%%%%%%%%%%%%%%%%
%%%%%%%%%%%%%%%%%%%%%%%%%%%%%%%%%%%%%%%%%%%%%%%%%%%%%%%%%%%%%%%%%%%%%

The power spectrum  is the variance of singleton fluctuations due to
quantum zero-point fluctuations. It  is easily defined by the
zero-point correlation function which could be computed when one
chooses   the Bunch-Davies vacuum state $|0\rangle$ in the
subhorizon limit.   The defining relation is given by
\begin{equation}
\langle0|\hat{\varphi}_a(\eta,0)\hat{\varphi}_b(\eta,0)|0\rangle=\int
\frac{dk}{k}{\cal P}_{ab},
\end{equation}
where $k=\sqrt{\bold{k}\cdot \bold{k}}$ is the comoving wave number.
Quantum fluctuations were created on all length scales with wave
number $k$. Cosmologically relevant fluctuations start their lives
inside the Hubble radius which defines the subhorizon: $k~\gg aH$.
On later, the comoving Hubble radius $1/(aH)$ shrinks during
inflation while keeping the wavenumber $k$ constant. Eventually, all
fluctuations exit the comoving Hubble radius, they reside on  the
superhorizon region of $k~\ll aH$ after horizon crossing.

For fluctuations of a massless scalar ($\bar{\nabla}^2\delta
\phi=0$) and tensor ($\bar{\nabla}^2h_{ij}=0$) with different
normalization originate on subhorizon scales and they propagate for
a long time on superhorizon scales. This can be checked by computing
their power spectra
\begin{eqnarray} \label{powerst}
{\cal P}_{\rm \delta\phi}&=&\frac{H^2}{(2\pi)^2}[1+z^2],\\
 \label{powerst1}{\cal P}_{\rm h}&=&2\times \Big(\frac{2}{M_{\rm
P}}\Big)^2\times{\cal P}_{\rm \phi}= \frac{2H^2}{\pi^2M^2_{\rm
P}}[1+z^2].
\end{eqnarray}

 To compute the singleton power spectrum, we have to know the commutation relations and the Wronskian condition. The canonical
conjugate momenta are given by
\begin{equation}
\pi_1=a^2\varphi'_2,~~\pi_2=a^2\varphi'_1.
\end{equation}
The canonical quantization is accomplished by imposing equal-time
commutation relations:
\begin{eqnarray}\label{comm}
[\hat{\varphi}_{1}(\eta,{\bf x}),\hat{\pi}_{1}(\eta,{\bf
y})]=i\delta^3({\bf x}-{\bf y}),~~[\hat{\varphi}_2(\eta,{\bf
x}),\hat{\pi}_{2}(\eta,{\bf y})]=i\delta^3({\bf x}-{\bf y}).
\end{eqnarray}
 The two operators $\hat{\varphi}_{1}$ and
$\hat{\varphi}_{2}$ are expanded in terms of Fourier modes
as~\cite{Mannheim:2004qz,Rivelles:2003jd,Myung:2015vya}
\begin{eqnarray}\label{hex1}
\hat{\varphi}_{1}(z,{\bf x})&=&\frac{1}{(2\pi)^{\frac{3}{2}}}\int
d^3{\bf k}N\Bigg[\Big(i\hat{a}_1({\bf k})\phi^1_{\bf k}(z)e^{i{\bf
k}\cdot{\bf
x}}\Big)+{\rm h.c.}\Bigg], \\
\label{hex2} \hat{\varphi}_2(z,{\bf
x})&=&\frac{1}{(2\pi)^{\frac{3}{2}}}\int d^3{\bf
k}\tilde{N}\Bigg[\Big(\hat{a}_2({\bf k})\phi^1_{\rm
k}(z)+\hat{a}_1({\bf k})\phi^2_{\rm k}(z)\Big)e^{i{\bf k}\cdot{\bf
x}}+{\rm h.c.}\Bigg]
\end{eqnarray}
with $N$ and $\tilde{N}$ the normalization constants.  Plugging
(\ref{hex1}) and (\ref{hex2}) into (\ref{comm}) determines the
relation of normalization constants as $N\tilde{N}=1/2k $ and
commutation relations between $\hat{a}_a({\bf k})$ and
$\hat{a}^{\dagger}_b({\bf k}')$ as
 \begin{equation} \label{scft}
 [\hat{a}_a({\bf k}), \hat{a}^{\dagger}_b({\bf k}')]= 2k
 \left(
  \begin{array}{cc}
   0 & -i  \\
    i & 1 \\
  \end{array}
 \right)\delta^3({\bf k}-{\bf k}'),
 \end{equation}
 where we observe a Jordan cell structure. This is the typical
 commutation relations appeared when one quantizes a degenerate Pais-Uhlenbeck
 fourth-order oscillator~\cite{Mannheim:2004qz}.
Here the commutation relation of $[\hat{a}_2({\bf k}),
\hat{a}^{\dagger}_2({\bf k}')]$ is implemented  by the Wronskian
condition. The Wronskian condition for $\phi^{1}_{{\bf k}}(z)$ and
$\phi^{2}_{{\bf k}}(z)$ leads to
\begin{eqnarray}
&&a^2\Big(\phi^1_{{\bf k}}\frac{d\phi^{2*}_{{\bf
k}}}{dz}-\phi^{2*}_{{\bf k}}\frac{d\phi^{1}_{{\bf
k}}}{dz}+\phi^{1*}_{{\bf k}}\frac{d\phi^{2}_{{\bf
k}}}{dz}-\phi^{2}_{{\bf k}}\frac{d\phi^{1*}_{{\bf
k}}}{dz}\Big)\nonumber
\\
&&=\sqrt{\frac{k}{2}}\frac{1}{H}\Big[2i(\tilde{c}_2-\tilde{c}^*_2)-(\tilde{c}_1+\tilde{c}_1^*)\Big(\frac{1}{z^3}+\frac{3}{z}\Big)\Big]~=~\frac{1}{k}.
\end{eqnarray}
To satisfy the above relation, let us  impose
\begin{eqnarray}
\tilde{c}_1=-\tilde{c}^*_1,~~~\tilde{c}_2=-\frac{iH}{2\sqrt{2k^3}}.
\end{eqnarray}

At this stage, it is worth to note  that the Wronskian normalization
condition was originally  designed for the second-order theory. In
the subhorizon limit of $z\to \infty$, the fourth-order contribution
is nothing, while it blows up unless $\tilde{c}_1$ is purely
imaginary in the superhorizon limit of $z\to 0$. Hence,  we may
neglect the fourth-order contribution to the Wronskian condition by
choosing $\tilde{c}_1$ to be purely imaginary.
 Considering  (\ref{c1mu}), one may
determine
\begin{eqnarray} \label{c1til} \tilde{c}_1=-i\frac{2H}{3\sqrt{2k^3}}
\end{eqnarray}
by choosing $\mu=2iH^2$. We note here  that choosing
$\tilde{c}_1=i\frac{2H}{3\sqrt{2k^3}}$ leads to the positive power
spectrum (${\cal P}_{22}>0$) in the whole range $z$, which
contradicts to the negative power spectrum of  a fourth-order scalar
theory.
%%%%%%%%%%%%%%%%%%%%%%%%%%%%%%%%%%%%%%%%%%%%%%%%%%%%%%%%%%%%%%%%%
\begin{figure}[t!]
\begin{center}
\begin{tabular}{cc}
\includegraphics[width=.8
\linewidth,origin=tl]{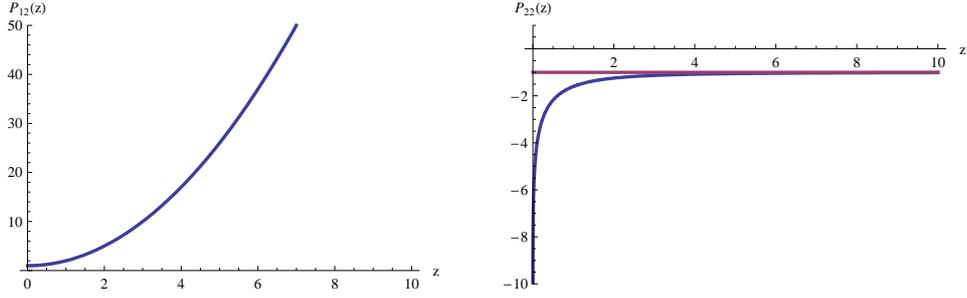}
\end{tabular}
\end{center}
\caption{Power spectra ${\cal P}_{12}$ and ${\cal P}_{22}$ as
functions of $z$ for $H^2=(2\pi)^2$. In the superhorizon  limit of
$z\to0$, one finds that ${\cal P}_{12} \to 1$ while ${\cal P}_{22}
\to -\infty$. On the other hand, ${\cal P}_{12}\to \infty$ and
${\cal P}_{22}\to-1$ in the subhorizon limit of $z\to \infty$. }
\end{figure}
%%%%%%%%%%%%%%%%%%%%%%%%%%%%%%%%%%%%%%%%%%%%%%%%%%%%%%%%%%%%%%%%%

Then, we could easily find that
\begin{equation} {\cal P}_{\rm 11}=0,~~{\cal P}_{\rm 12}(z)={\cal
P}_{\rm 21}(z)=\frac{k^3}{2\pi^2}|\phi_{\bf
k}^1|^2=\frac{H^2}{(2\pi)^2}[1+z^2].
\end{equation}
However, the power spectrum ${\cal P}_{\rm 22}$ takes a complicated
form
\begin{eqnarray}\label{pw22}
{\cal P}_{\rm 22}(z)&\equiv& {\cal P}_{\rm 22}^{(1)}(z)+{\cal
P}_{\rm
22}^{(2)}(z)\nonumber\\
&=&\frac{k^3}{2\pi^2}\Big[|\phi_{\bf k}^1|^2+i(\phi_{\bf
k}^1\phi_{\bf k}^{2*}-\phi_{\bf k}^2\phi_{\bf k}^{1*})\Big]\nonumber\\
&=&\left(\frac{H}{2\pi}\right)^2\Big[1+z^2
-\Big\{1+z^2+4i\tilde{c}_1\frac{\sqrt{2k^3}}{H}+2i\tilde{c}_1\frac{\sqrt{2k^3}}{H}{\bf
Re}[f(z)]\Big\}\Big]\nonumber\\
 &=&-\frac{4}{3}\left(\frac{H}{2\pi}\right)^2\Big[2+{\bf
Re}[f(z)]\Big],
\end{eqnarray}
where $f(z)$ is given by
\begin{eqnarray}
f(z)=e^{2iz}(i+z)^2{\rm Ei}(-2iz).
\end{eqnarray}
This is another  of our main results: power spectrum of massless
singleton is explicitly expressed in terms of the
exponential-integral function. Fig. 3 indicates the behaviors of
${\cal P}_{12}(z)$ and ${\cal P}_{22}(z)$ generated during dS
inflation. We note that the former shows a typical power spectrum
for a massless scalar ($\delta \phi,\varphi_1$) or graviton ($h$),
while the latter indicates a power spectrum of the singleton
($\varphi_2$). It is reasonable to assist  that the power spectrum
of ${\cal P}_{22}$ is negative because it corresponds to that of a
purely fourth-order  scalar theory. That is, one could not avoid to
find ghost-instability when computing the power spectrum of a
fourth-order derivative scalar theory during dS inflation.

In the subhorizon limit of $z\to \infty$, one finds a negatively
scale-invariant spectrum
\begin{eqnarray}\label{pw23sub}
{\cal P}^{z\to\infty}_{\rm 22}&=&-\left(\frac{H}{2\pi}\right)^2
\end{eqnarray}
because ${\bf Re}[f(z)]\to -\frac{5}{4}$ in this limit. We note that
the power spectrum  of the scale-invariant scalar tensor theory is
given by~\cite{Myung:2015hua}
\begin{equation}\label{pw23sec}
{\cal P}_{\rm SIST} =\frac{1}{2(2\pi)^2}.
\end{equation}
 Taking  $f(z)$ in
the superhorizon limit of $z\to0$
\begin{eqnarray}
f(z)\to_{z\to 0}
\Big[-\gamma-\ln[2z]+(1-\gamma)z^2\Big]+i\Big[-\frac{\pi}{2}+2z-\frac{\pi
z^2}{2}\Big],
\end{eqnarray}
the power spectrum (\ref{pw22}) of massless singleton leads to
\begin{eqnarray}\label{pw24}
{\cal P}^{z\to0}_{\rm
22}(z)&=&-\frac{4}{3}\left(\frac{H}{2\pi}\right)^2\Big(2-\gamma-\ln[2z]\Big)
\\
&=&\frac{4}{3}\left(\frac{H}{2\pi}\right)^2\Big(\ln[z]-0.73\Big),
\nonumber
\end{eqnarray}
which  explains why ${\cal P}_{22}(z)$ blows up negatively  as $z\to
0$ in Fig. 3. On the other hand, one has a  power spectrum for a
massless scalar
\begin{eqnarray}\label{pw25}
{\cal P}^{z\to0}_{\rm 12}&=&\left(\frac{H}{2\pi}\right)^2.
\end{eqnarray}
%%%%%%%%%%%%%%%%%%%%%%%%%%%%%%%%%%%%%%%%%%%%%%%%%%%%%%%%%%%%%%%%%%%%%
%%%%%%%%%%%%%%%%%%%%%%%%%%%%%%%%%%%%%%%%%%%%%%%%%%%%%%%%%%%%%%%%%%%%%
%%%%%%%%%%%%%%%%%%%%%%%%%%%%%%%%%%%%%%%%%%%%%%%%%%%%%%%%%%%%%%%%%%%%%
\section{Discussions}

We have obtained the exact solution and  computed the complete
power spectrum (\ref{pw22}) of a
 singleton expressed in term of the exponential-integral function  by solving
the degenerate fourth-order equation and by requiring the
Pais-Uhlenbeck quantization scheme for a degenerate fourth-order
oscillator.

Its two asymptotic behaviors are quite different from those
[(\ref{pw23sec}) and (\ref{pw25})] of a massless scalar. In the
subhorizon limit $z\to \infty$, the power spectrum (\ref{pw23sub})
of a singleton  is a negatively scale-invariant one which is
opposite to (\ref{pw23sec})  of scale-invariant scalar-tensor
theory~\cite{Myung:2015hua}, while it blows up (negatively
divergent) in the superhorizon limit of $z\to 0$ as is shown in
(\ref{pw24}). This  indicates a feature of purely fourth-order
derivative scalar theory in dS spacetime~\cite{Kehagias:2012pd}.

Even though our computation was based on the dS inflation, the above
asymptotic features have suggested that the Einstein-singleton
theory including  a fourth-order scalar theory  is not a good
candidate for a slow-roll (quasi-dS) inflation model.

Finally, we discuss some issues relevant  to our model. \\
$\bullet$  Ghost-instability of the model\\
Since $S_{\rm S}$ in (\ref{SGA}) reduces to the fourth-order
derivative  scalar theory (\ref{singleton}), we worry about the
ghost-instability problem. Using the Pais-Uhlenbeck quantization
scheme for a degenerate fourth-order oscillator in dS spacetime, we
have found the negative power spectrum ${\cal P}_{22}(z)$ in
(\ref{pw22}), depicted in Fig 3. In the subhorizon limit of $z\to
\infty$, we have obtained a negatively scale-invariant power
spectrum (\ref{pw23sub}) which indicates  the ghost instability
clearly. On the other hand, ${\cal P}_{22}(z)$ blows negatively up
in the superhorizon limit of $z\to 0$. This  indicates that the
singleton theory is a fourth-order
derivative scalar theory which must  contain a ghost state. \\
$\bullet$  Problem of exit mechanism\\The dS inflation is driven by
the cosmological constant $\Lambda$ which is a non-dynamical
quantity. Hence, this corresponds to  an eternal inflation and thus,
there is no natural way to exit the inflationary phase. This is a
handicap of dS inflation. In the slow-roll inflation (quasi-dS
inflation), however, the inflaton plays an essential role in exiting
the inflationary
phase.  \\
$\bullet$  Is $\mu=2iH^2$  a mass square of $\phi_1$?\\
In order to obtain Eq. (\ref{c1til}), we specified $\mu=2iH^2$.
Recalling the definition of $\mu$ in (\ref{SGA}), it seems that
$\mu$ plays the role of the mass square of $\phi_1$. However, this
is not true. $\mu$ is just a parameter of connecting $\phi_2$ with
$\phi_1$ to get the fourth-order derivative equation for $\phi_2$
from a mixed kinetic term. If one wishes to have a massive
singleton, one has to include a potential term of
$m^2\phi_1\phi_2$~\cite{Myung:2014pza}:
$(\bar{\nabla}^2-m^2)\varphi_1=0,~(\bar{\nabla}^2-m^2)\varphi_2=\mu\varphi_1,~
(\bar{\nabla}^2-m^2)^2\varphi_2=0$. \\
$\bullet$ Slow-roll inflation in the Einstein-singleton theory\\If
one wishes to consider  the slow-roll inflation in the
Einstein-singleton theory $S_{\rm ES}$ including the potential of $
m^2\phi_1\phi_2$, the Einstein equation takes the form of
$G_{\mu\nu}=T^m_{\mu\nu}/M^2_{\rm P}$ which provides the energy
density
$\rho=\dot{\phi_1}\dot{\phi_2}+(m^2\phi_1\phi_2+\mu\phi_1^2/2)$ and
the pressure
$p=\dot{\phi_1}\dot{\phi_2}-(m^2\phi_1\phi_2+\mu\phi_1^2/2)$. The
first and second Friedmann equations are given by
$H^2=\frac{\rho}{3M^2_{\rm P}}$ and
$\dot{H}=-\frac{\rho+p}{2M^2_{\rm P}}$. Even though  this model is
similar to two-field inflation model with the chaotic potentials,
this is not the case because their  full scalar equations are given
by $\ddot{\phi}_1+3H(t)\dot{\phi}_1+m^2\phi_1=0$ and
$\ddot{\phi}_2+3H(t)\dot{\phi}_2+m^2\phi_2=-\mu\phi_1$ which are
combined to give a fourth-order equation  of
$(\frac{d^2}{dt^2}+3H(t)\frac{d}{dt}+m^2)^2\phi_2=0$. It conjectures
that their slow-roll equations are quite different from those of
two-field inflation. Furthermore, it requires a non-trivial task to
perform the cosmological perturbations around the slow-roll
inflation instead of the dS inflation. Especially, it is important
to define the curvature perturbation ${\cal R}$ in the
Einstein-singleton theory. It was  given by ${\cal R}=-H\delta
\phi/\dot{\phi}$ for  the single-field inflation in spatially flat
gauge, while it takes the form of ${\cal R}_{\rm
S}=-H[\varphi_1/\dot{\phi_1}+\varphi_2/\dot{\phi_2}]$ for the
singleton inflation. For example, the power spectrum appeared in dS
spacetime with
$\dot{\phi}_1=\dot{\phi}_2=0$~\cite{Kehagias:2012pd,Myung:2014pza}
was given by ${\cal P}^m_{\varphi_2\varphi_2} \sim
z^{2w}(1+2\ln[z])$ with $w=3/2-\sqrt{9/4-m^2/H^2}$ in the
superhorizon limit.
 However, we
remain ``cosmological perturbations of the Einstein-singleton theory
around the slow-roll inflation" as a future work, worrying about the
appearance of the ghost states. This is so because the strange
asymptotic behavior of power spectrum of ${\cal
P}_{\varphi_2\varphi_2}$ indicates a negatively divergent behavior
in the superhorizon limit of $z\to 0$, which reflects that the
Einstein-singleton theory includes a fourth-order derivative scalar
theory.  Furthermore, there is no way to avoid a ghost-instability
in the whole range of $z$.  Thus, our result during dS inflation
suggests  that the Einstein-singleton theory is not considered as a
model for developing a slow-roll inflation because a negative power
spectrum of curvature perturbation (${\cal P}_{{\cal R}_{\rm S}{\cal
R}_{\rm S}}<0$) persists in the slow-roll inflation. This is because
$\varphi_2$ satisfies a fourth-order differential  equation during
the slow-roll inflation.

%%%%%%%%%%%%%%%%%%%%%%%%%%%%%%%%%%%%%%%%%%%%%%%%%%%%%%%%%%%%%%%%%%%%%
%%%%%%%%%%%%%%%%%%%%%%%%%%%%%%%%%%%%%%%%%%%%%%%%%%%%%%%%%%%%%%%%%%%%%
%%%%%%%%%%%%%%%%%%%%%%%%%%%%%%%%%%%%%%%%%%%%%%%%%%%%%%%%%%%%%%%%%%%%%


\newpage


\begin{thebibliography}{99}

%\cite{Ade:2015lrj}
\bibitem{Ade:2015lrj}
  P.~A.~R.~Ade {\it et al.} [Planck Collaboration],
  %``Planck 2015 results. XX. Constraints on inflation,''
  arXiv:1502.02114 [astro-ph.CO].
  %%CITATION = ARXIV:1502.02114;%%
   %400 citations counted in INSPIRE as of 26 Nov 2015

%\cite{Starobinsky:1980te}
\bibitem{Starobinsky:1980te}
  A.~A.~Starobinsky,
  %``A New Type of Isotropic Cosmological Models Without Singularity,''
  Phys.\ Lett.\ B {\bf 91}, 99 (1980).  doi:10.1016/0370-2693(80)90670-X
   %%CITATION = doi:10.1016/0370-2693(80)90670-X;%%
    %2518 citations counted in INSPIRE as of 26 Nov 2015


%\cite{Kehagias:2012pd}
\bibitem{Kehagias:2012pd}
  A.~Kehagias and A.~Riotto,
  %``Operator Product Expansion of Inflationary Correlators and Conformal Symmetry of de Sitter,''
  Nucl.\ Phys.\ B {\bf 864}, 492 (2012)  doi:10.1016/j.nuclphysb.2012.07.004  [arXiv:1205.1523 [hep-th]].
  %%CITATION = doi:10.1016/j.nuclphysb.2012.07.004;%%
  %70 citations counted in INSPIRE as of 26 Nov 2015

%\cite{Myung:2014pza}
\bibitem{Myung:2014pza}
  Y.~S.~Myung and T.~Moon,
  %``Cosmological singleton gravity theory and dS/LCFT correspondence,''
  JHEP {\bf 1410}, 137 (2014)  doi:10.1007/JHEP10(2014)137  [arXiv:1407.7742 [gr-qc]].
  %%CITATION = doi:10.1007/JHEP10(2014)137;%%
  %1 citations counted in INSPIRE as of 26 Nov 2015


%\cite{Pais:1950za}
\bibitem{Pais:1950za}
  A.~Pais and G.~E.~Uhlenbeck,
  %``On Field theories with nonlocalized action,''
  Phys.\ Rev.\  {\bf 79}, 145 (1950).
  %%CITATION = PHRVA,79,145;%%
   %314 citations counted in INSPIRE as of 15 Jul 2014

%\cite{Mannheim:2004qz}
\bibitem{Mannheim:2004qz}
  P.~D.~Mannheim and A.~Davidson,
  %``Dirac quantization of the Pais-Uhlenbeck fourth order oscillator,''
  Phys.\ Rev.\ A {\bf 71}, 042110 (2005)  [hep-th/0408104].
   %%CITATION = HEP-TH/0408104;%%  %35 citations counted in INSPIRE as of 15 Jul 2014

%\cite{Kim:2013mfa}
\bibitem{Kim:2013mfa}
  Y.~W.~Kim, Y.~S.~Myung and Y.~J.~Park,
  %``Quantization of n coupled scalar field theory,''
   Phys.\ Rev.\ D {\bf 88}, 085032 (2013)  doi:10.1103/PhysRevD.88.085032  [arXiv:1307.6932].
    %%CITATION = doi:10.1103/PhysRevD.88.085032;%%
    %6 citations counted in INSPIRE as of 02 Dec 2015



%\cite{Flato:1986uh}
\bibitem{Flato:1986uh}
  M.~Flato and C.~Fronsdal,
  %``The Singleton Dipole,''
  Commun.\ Math.\ Phys.\  {\bf 108}, 469 (1987).
  %%CITATION = CMPHA,108,469;%%
   %23 citations counted in INSPIRE as of 15 Jul 2014

%\cite{Ghezelbash:1998rj}
\bibitem{Ghezelbash:1998rj}
  A.~M.~Ghezelbash, M.~Khorrami and A.~Aghamohammadi,
  %``Logarithmic conformal field theories and AdS correspondence,''
  Int.\ J.\ Mod.\ Phys.\ A {\bf 14}, 2581 (1999)  [hep-th/9807034].
  %%CITATION = HEP-TH/9807034;%%  %57 citations counted in INSPIRE as of 15 Jul 2014



%\cite{Kogan:1999bn}
\bibitem{Kogan:1999bn}
  I.~I.~Kogan,
  %``Singletons and logarithmic CFT in AdS / CFT correspondence,''
  Phys.\ Lett.\ B {\bf 458}, 66 (1999)  [hep-th/9903162].
  %%CITATION = HEP-TH/9903162;%%
  %52 citations counted in INSPIRE as of 15 Jul 2014

%\cite{Myung:1999nd}
\bibitem{Myung:1999nd}
  Y.~S.~Myung and H.~W.~Lee,
  %``Gauge bosons and the AdS(3) / LCFT(2) correspondence,''
  JHEP {\bf 9910}, 009 (1999)  [hep-th/9904056].
  %%CITATION = HEP-TH/9904056;%%
   %43 citations counted in INSPIRE as of 15 Jul 2014

%\cite{Grumiller:2013at}
\bibitem{Grumiller:2013at}
  D.~Grumiller, W.~Riedler, J.~Rosseel and T.~Zojer,
  %``Holographic applications of logarithmic conformal field theories,''
  J.\ Phys.\ A {\bf 46}, 494002 (2013)  [arXiv:1302.0280 [hep-th]].
  %%CITATION = ARXIV:1302.0280;%%
  %19 citations counted in INSPIRE as of 15 Jul 2014

%\cite{Rivelles:2003jd}
\bibitem{Rivelles:2003jd}
  V.~O.~Rivelles,
  %``Triviality of higher derivative theories,''
  Phys.\ Lett.\ B {\bf 577}, 137 (2003)  doi:10.1016/j.physletb.2003.10.039  [hep-th/0304073].
   %%CITATION = doi:10.1016/j.physletb.2003.10.039;%%
   %24 citations counted in INSPIRE as of 26 Nov 2015

\bibitem{AS}
M. Abramowitz and A. Stegun, Handbook of Mathematical functions,
(Dover publications, New York, 1970).


%\cite{Myung:2015vya}
\bibitem{Myung:2015vya}
  Y.~S.~Myung and T.~Moon,
  %``Scale-invariant tensor spectrum from conformal gravity,''
  Mod.\ Phys.\ Lett.\ A {\bf 30}, no. 32, 1550172 (2015)  doi:10.1142/S0217732315501722  [arXiv:1501.01749 [gr-qc]].
  %%CITATION = doi:10.1142/S0217732315501722;%%
  %2 citations counted in INSPIRE as of 26 Nov 2015

%\cite{Myung:2015hua}
\bibitem{Myung:2015hua}
  Y.~S.~Myung and Y.~J.~Park,
  %``Scale-invariant power spectra from a Weyl-invariant scalar?tensor theory,''
   Eur.\ Phys.\ J.\ C {\bf 76}, no. 2, 79 (2016)  doi:10.1140/epjc/s10052-016-3924-0  [arXiv:1508.04188 [gr-qc]].
    %%CITATION = doi:10.1140/epjc/s10052-016-3924-0;%%  %1 citations counted in INSPIRE as of 12 May 2016




\end{thebibliography}
\end{document}